# Quardratic Electromechanical Strain in Silicon Investigated by Scanning Probe Microscopy

Junxi Yu [1,2,a], Ehsan Nasr Esfahani [3,a], Qingfeng Zhu [1], Dongliang Shan [1], Tingting Jia [2], Shuhong Xie [1,b], Jiangyu Li [2,3,b]

1. Key Laboratory of Low Dimensional Materials and Application Technology of Ministry of Education, and School of Materials Science and Engineering, Xiangtan University, Xiangtan, Hunan, 411105, China
2. Shenzhen Key Laboratory of Nanobiomechanics, Shenzhen Institutes of Advanced Technology, Chinese Academy of Sciences, Shenzhen, Guangdong, 518055, China
3. Department of Mechanical Engineering, University of Washington, Seattle, WA, 98195, USA

## Abstract

Piezoresponse force microscopy (PFM) is a powerful tool widely used to characterize piezoelectricity and ferroelectricity at the nanoscale. However, it is necessary to distinguish microscopic mechanisms between piezoelectricity and non-piezoelectric contributions measured by PFM. In this work, we systematically investigate the first and second harmonic apparent piezoresponses of silicon wafer in both vertical and lateral modes, and we show that it exhibits apparent electromechanical response that is quadratic to the applied electric field, possibly arising from ionic electrochemical dipoles induced by the charged probe. As a result, the electromechanical response measured is dominated by the second harmonic response in vertical mode, and its polarity can be switched by the DC voltage with evolving coercive field and maximum amplitude, in sharp contrast with typical ferroelectric materials we used as control. The ionic activity in silicon is also confirmed by scanning thermo-ionic microscopy (STIM) measurement, and this work points toward a set of methods to distinguish true piezoelectricity from the apparent ones.

---

# 1. Introduction

In the last 20 years, piezoresponse force microscopy (PFM) has emerged as one of the most powerful tools to study piezoelectricity and ferroelectricity at the nanoscale [1–5], yielding considerable insight into domains, defects, nucleation and switching of ferroelectric materials. Despite its successes, it has been increasingly recognized by the community that the apparent piezoresponse signal in PFM may arise from a number of distinct microscopic mechanisms other than piezoelectric effect [6–10], and thus it may be artifacts that do not reflect the true piezoelectricity and ferroelectricity of the materials probed. This is particularly problematic for silicon wafers and silica glasses, two commonly used substrates that are not piezoelectric, yet both of them are found to exhibit not only apparent piezoresponse when probed by PFM, but also polarity switching under DC bias [6,7]. It is thus important to understand the characteristics and mechanism of such apparent piezoresponse in order to guide the experimental design as well as data analysis and interpretation of PFM for probing true piezoelectric and ferroelectric behaviors.

Chen *et al.* has discussed various microscopic mechanisms that contribute to apparent piezoresponse signal [8], including electrostrictive strains, electrochemical dipoles, electrostatic interactions, and ionic Vegard strain. In this work, we expand this analysis and apply it to investigate the apparent piezoresponse observed in silicon wafer in details. We show that the apparent piezoresponse in silicon is dominated by nonlinear strain quadratic to the applied voltage, and thus may arise from ionic electrochemical dipoles induced by the charged probe. This is supported by the observed switching of piezoresponse polarity under DC bias, which can be explained by the reversed electrochemical dipoles, and the strain is found to increase with DC bias as a result. Systematic comparisons have been made with classical piezoelectric and ferroelectric materials to highlight their different characteristics, and the recently developed scanning thermo-ionic microscopy (STIM) has been used to further confirm the ionic origin of such response [11,12]. The study thus sheds insight into the microscopic mechanisms as well as characteristics of apparent piezoresponse observed in silicon. In order to help the community to analyze such problems in other materials and systems, we also make our software codes used for data acquisition and analysis publicly accessible online [13].

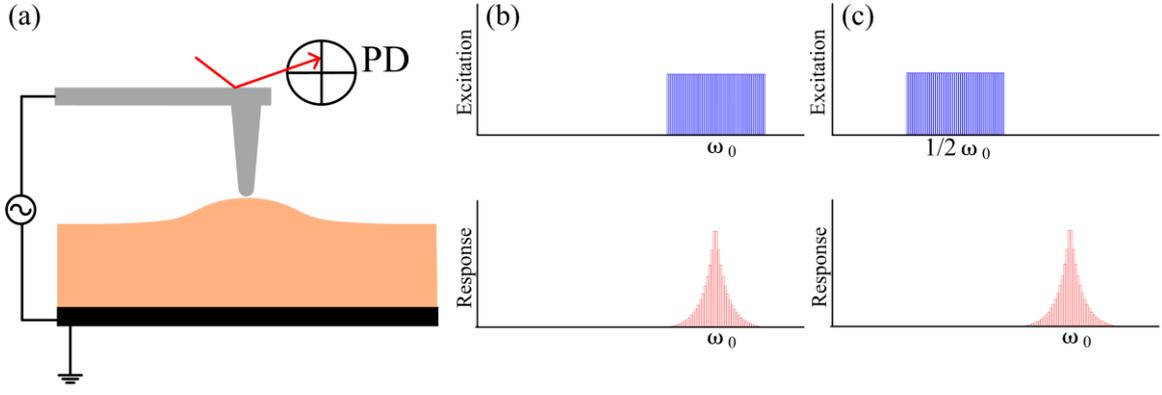

**Fig. 1** Schematics of PFM setup; (a) dynamic strain induced by the charged probe measured by photo-diode; (b) first harmonic piezoresponse; and (c) second harmonic piezoresponse.

## 2. Methods
### 2.1 Principle

Any dielectric would be deformed by an electric field, which is known as electrostrictive effect with electrostrictive strain given by $\varepsilon = QP^2$, where $Q$ is the electrostrictive coefficient that is usually very small, whereas $P$ is polarization that consists of contributions induced by electric field $E$ through dielectric susceptibility $\chi$ and spontaneous polarization $P_s$ if any, $P = \chi E + P_s$. Therefore, if we apply an AC electric field on top of a DC one to the dielectric, $E = E_0 + E_a e^{i\omega t}$, where $\omega$ is the angular frequency, then the strain resulted is given by

$$\varepsilon = Q[\chi^2(E_0^2 + E_a^2 e^{i2\omega t} + 2E_0 E_a e^{i\omega t}) + P_s^2 + 2\chi(E_0 + E_a e^{i\omega t})P_s]. \tag{1}$$

In typical PFM experiments, it is the dynamic strain that is measured near the probe-sample contact resonance frequency through lock-in amplifier to enhance the sensitivity, as schematically shown in Fig. 1. As such, we discard the static strain, and focus on the first and second harmonics of the strain responses,

$$\varepsilon = 2Q\chi E_a (\chi E_0 + P_s) e^{i\omega t} + Q\chi^2 E_a^2 e^{i2\omega t}. \tag{2}$$

A number of observations can be made from Eq. (2). First of all, when there is relatively large spontaneous polarization, as in a typical ferroelectric, then the strain response is predominantly reflected by the first harmonic with its magnitude linear to the electric field, whereas second harmonic response is much smaller. In addition, if the spontaneous polarization is switched by a DC electric field, then the polarity of the first harmonic strain is switched as well, and the saturated strain response is insensitive to the DC magnitude beyond switching. This is the basis of PFM. However, if there is no spontaneous polarization, and in the absence of a DC field, then the second harmonic strain response will dominate the first harmonic one and its magnitude should be quadratic to the applied electric field. Thus it is possible to distinguish piezoelectric strain from non-piezoelectric one by comparing the first and second harmonic responses, as suggested by Chen *et al*. [8], which can be implemented as shown in Fig. 1. The first harmonic response can be measured by exciting the sample and measuring the response near the probe-sample contact resonance $\omega_0$ (Fig. 1b), whereas second harmonic response can be measured by exciting the sample near $\omega_0/2$ and measuring the corresponding response near $\omega_0$ (Fig. 1c).

## 2.2 Implementation

In order to compare and analyze the characteristics of piezoresponse of different types of materials, we investigated p-type single crystalline silicon (100) wafer of resistivity 5~10 Ω·cm obtained from Hangzhou Jingbo Technology Co., as well as ferroelectric samples including bismuth ferrite (BFO), periodically poled lithium niobate (PPLN), and lead zirconate titanate (PZT). Both of our PFM and STIM experiments utilized Asylum Research MFP-3D-Bio AFM in combination with a Zurich Instrument HF2LI lock-in amplifier and proportional-integral-derivative (PID) controller. In PFM measurements, an ASYELEC-01 conductive probe coated with Pt/Ir having a spring constant of 2 N/m and resonance frequency in air around 70 kHz was used. In STIM measurements, an Anasys ThermaLever AN2-300 scanning thermal probe with a micro-fabricated heater integrated at the end of the cantilever having a spring constant of 0.2-0.5 N/m and resonance frequency in air around 30 kHz was used.

First and second harmonic PFM measurements were first carried out by recording amplitude responses of the cantilever as a function of frequency under an AC voltage of 20 V (for Si) and 9 V (for BFO), and then repeated with AC voltage between 10 and 20 V for Si and between 2 and 12 V for BFO, both having 1 V increment. A code was developed for data acquisition and analysis of first and second harmonic responses versus AC excitation voltage in both vertical and lateral PFM at selected points [13], which evaluates resonance frequency, quality factor, as well as amplitude and phase at resonance based on simple harmonic oscillator (SHO) model [14].

For polarity switching, a sequence of DC voltages with triangular step function imposed by the AC voltage was applied to the sample surface through the conductive probe, and the phase and amplitude response were measured during the "Off" state when DC voltage was stepped back to zero to minimize the electrostatic interactions [15]. Two cycles of such spectroscopic measurements were carried out on each point of Si and PZT with AC voltage being 10 V and 0.8 V, respectively.

For PFM mappings, 200 × 200 nm$^2$ and 11× 11 μm$^2$ areas of Si and PPLN were scanned via dual amplitude resonance tracking (DART) technique [16], using HF2LI lock-in amplifier and PID controller integrated with Asylum Research MFP-3D-Bio AFM. The bimodal drive signal of the conductive probe comes from the output of the lock-in, and the deflection signal of the probe is received by the input of the lock-in, wherein the PID controller is used to regulate the difference between detected amplitudes and adjust the drive frequency.

For measuring the second and fourth harmonic STIM responses of Si, 4 V AC heating voltage was applied to the thermal probe [11,12] near $\omega_0/2$ and $\omega_0/4$, respectively, and the amplitude responses of the cantilever were measured near the cantilever-sample contact resonant frequency $\omega_0$ through Zurich lock-in amplifier. The heating voltage induces localized temperature rise and thus thermal expansion that are second harmonic to the heating frequency through resistive heating, and the resulted thermal stress induces additional ionic fluctuation and thus Vegard strain that are fourth harmonic to the heating frequency. As such, by measuring second and forth harmonic STIM responses, information on local thermal expansion and ionic Vegard strain can be obtained [11,12]. In addition, the second and fourth harmonic STIM responses of the cantilever as a function of heating voltage at resonance frequency were also measured by applying 1-5 V heating voltage with 1 V increment to the thermal probe. For STIM mappings, 200 × 200 nm$^2$ area of Si was scanned by MFP-3D-Bio AFM using HF2LI lock-in amplifier and PID controller under DART [16].

## 3. Results and Discussions

We first study the first and second harmonic piezoresponses of Si wafer in comparison with BFO film, a typical ferroelectric material, as shown in Fig. 2. Both vertical and lateral piezoresponses are examined versus the excitation frequency, and typical results are given in Fig. 2ab showing evident contrast. As predicted by Eq. (2), Si wafer is observed to have much higher second harmonic response than the first harmonic one, measured near the vertical and lateral resonant frequencies of 288 kHz and 790 kHz, respectively, due to its lack of spontaneous polarization and intrinsic piezoelectricity, whereas the trend is reversed for BFO, with the first harmonic response much higher than the second harmonic one due to its ferroelectricity, measured near the vertical and lateral resonant frequencies of 274 kHz and 783 kHz, respectively. The higher resonant frequency of Si also indicates its higher stiffness [17,18]. It is also interesting to note that Si has a non-negligible first harmonic response in the absence of DC voltage, which may arise from ionic Vegard strain [19,20]. This suggests that the second harmonic strain measured may be induced by ionic electrochemical dipoles according to Eq. (2), possibly arising from electrochemical oxidation of silicon [21] and the resulted interfacial dipoles [22], though electrostatic contribution cannot be excluded as well. If this is the case, then the lateral piezoresponse of Si should be negligibly small due to the high symmetry of induced dipoles, which is indeed observed in Fig. 2a for both first and second harmonic responses in lateral mode, whereas BFO has significant first harmonic lateral response that is much higher than second harmonic one (Fig. 2b). Even more importantly, the intrinsic piezoresponses are calculated from resonant peaks using simple harmonic oscillator (SHO) model [14], and the vertical amplitudes are plotted versus AC voltage as shown in Fig. 2cd, averaged over 10 spatial points. The contrast seen in Fig. 2ab is confirmed not only throughout the range of voltage applied, but also consistent for the different spatial locations probed as indicated by the small error bars. In addition, while the first harmonic lateral responses of BFO (Fig. 2f) exhibit a clear linear behavior with increased AC voltage (except at higher voltage end), lateral responses of Si versus the AC voltage, as is shown in Fig. 2e, do not reveal any clear trend, suggesting that the responses are probably just background noises.

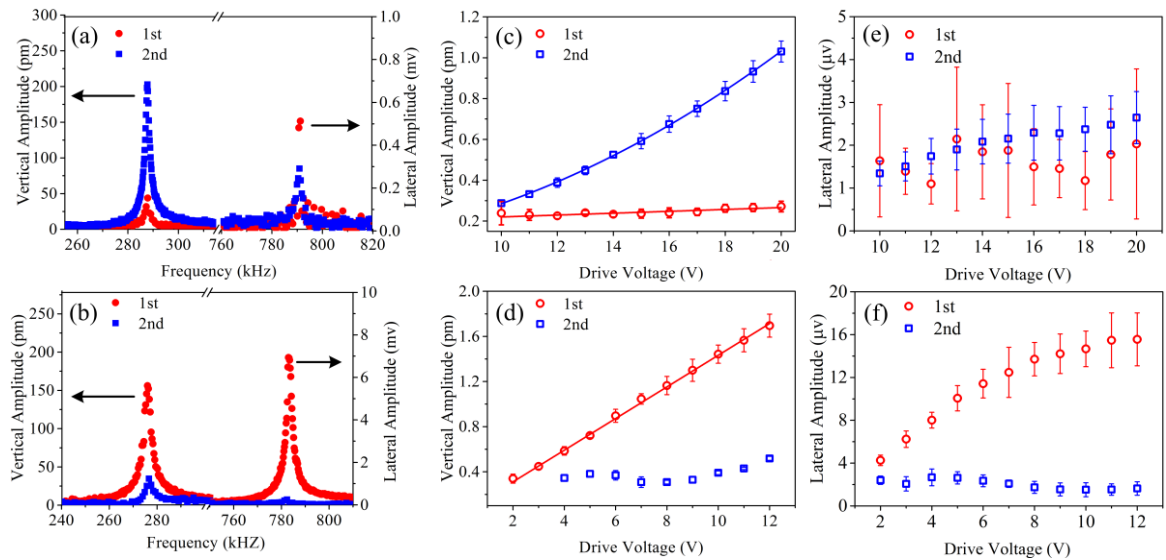

**Fig. 2** Comparison of the first and second harmonic responses of Si wafer and BFO film; (ab) vertical and lateral piezoresponses of (a) Si and (b) BFO versus excitation frequency; (cd) vertical amplitude of (c) Si and (d) BFO versus AC voltage; (ef) lateral amplitude of (e) Si and (f) BFO versus AC voltage.

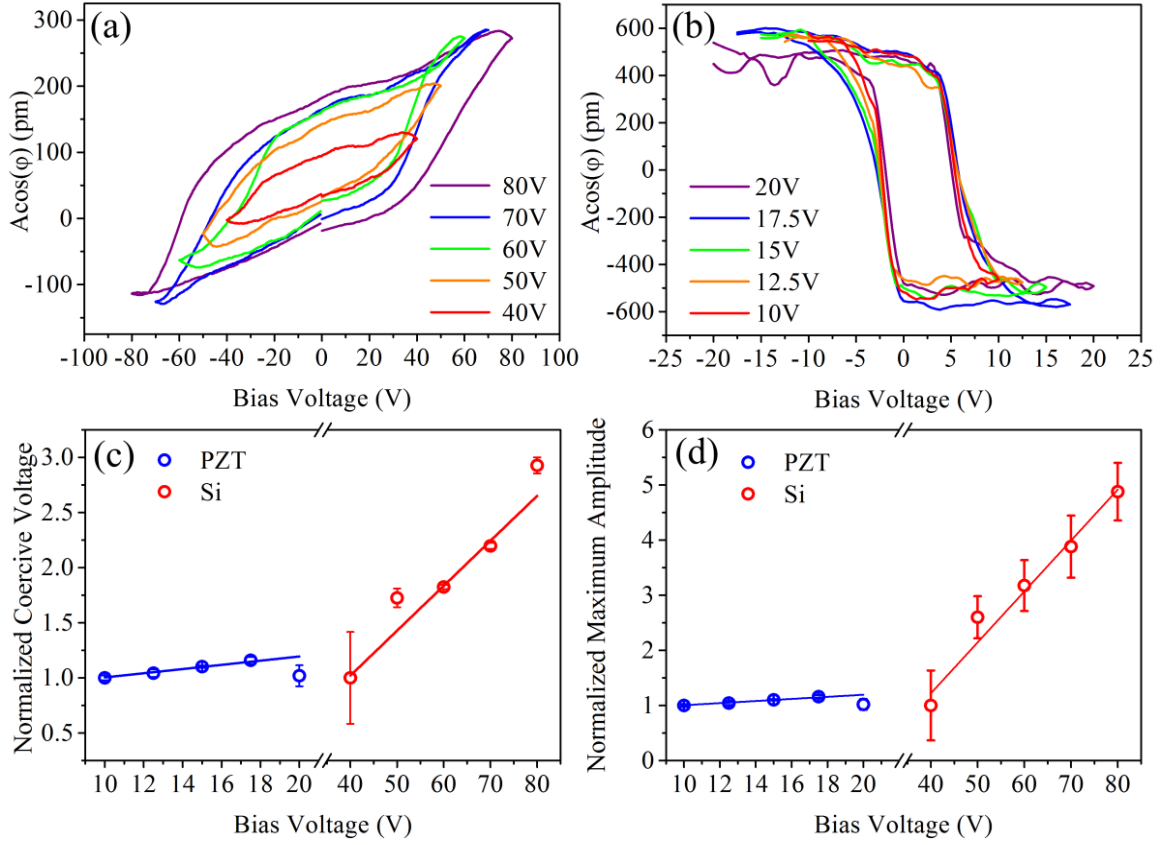

**Fig. 3** Comparison of switching behavior of Si and PZT; (ab) hysteresis loops of (a) Si and (b) PZT; (c) normalized coercive voltage versus DC bias; (d) normalized maximum piezoresponse amplitude versus DC bias. (Maximum piezoresponse is define as half of the height of the loops and coercive voltage is define as half of the width of the loops).

The dominance of second harmonic piezoresponse in Si suggests that it may originate from electrochemical dipoles, and if this is the case, then it is possible to switch the polarity of the first harmonic piezoresponse. This is indeed what we observed in silicon, wherein a sequence of cycling DC biases are applied to trigger possible polarity switch, while AC voltage is simultaneously applied to excite the piezoresponse that is measured during the OFF state when the DC bias is stepped back to zero [15], resulting in characteristic phase-voltage ($\varphi$-$V$) hysteresis loop and amplitude-voltage ($A$-$V$) butterfly loop that are combined as $A\cos\varphi$-$V$ hysteresis loops, as shown in Fig. 3ab for Si and PZT thin film. While both materials show clear polarity switching under the DC voltage, their characteristics are quite different. PZT, a classical ferroelectric material, has a distinct coercive voltage that remains stable with increased maximum DC voltage, and the saturated piezoresponse amplitude is stable as well. Si, on the other hand, has a nominal coercive voltage that increases with the maximum DC voltage applied, so does the maximum piezoresponse amplitude, which does not appear to saturate either. The contrast can be more clearly seen in Fig. 3cd, wherein normalized coercive voltage and maximum piezoresponse amplitude versus DC bias are compared between Si and PZT. The difference can be explained on the basis of spontaneous polarization in ferroelectrics that is independent of external electric field, as in PZT, so that both coercive voltage and maximum piezoresponse are insensitive to the maximum DC bias applied. The electrochemical dipole moment induced by an external electric field, on the other hand, increases with the strength of the electric field applied, and as a result, larger coercive voltage would be needed to switch it, and higher apparent piezoresponse is resulted at higher DC bias. The observation thus further supports that the apparent piezoresponse in Si arises from induced electrochemical dipoles.

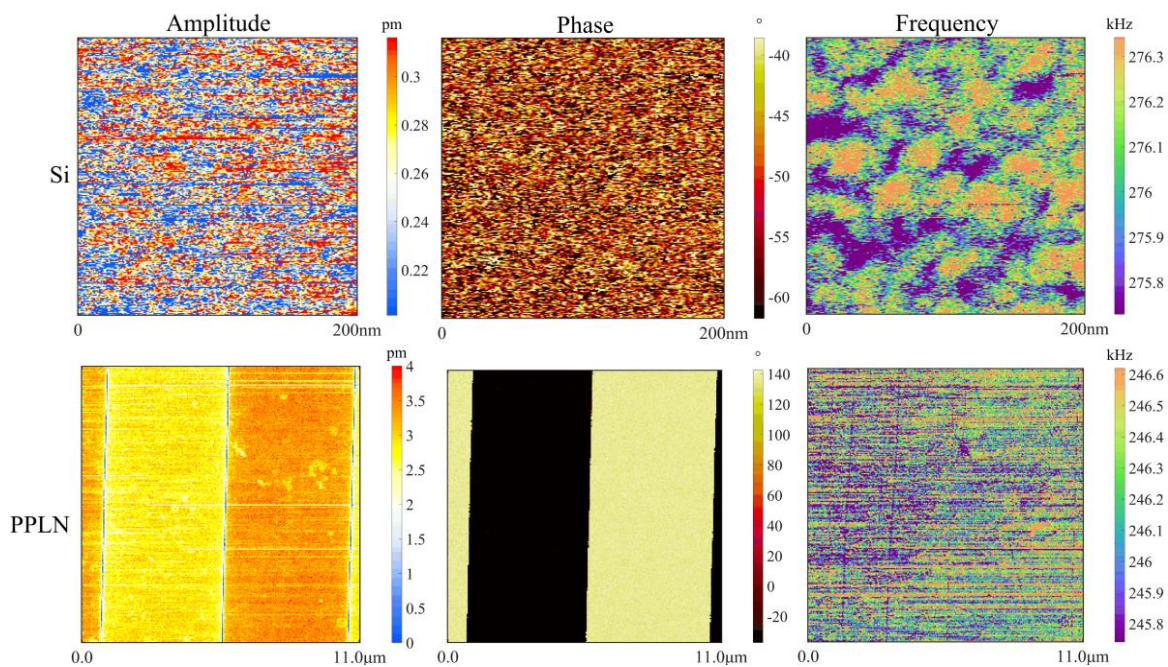

**Fig. 4** Comparison of first harmonic piezoresponse mappings between Si and LiNbO$_3$, including amplitude, phase and frequency.

Another implication of induced electrochemical dipole is that its polarity, i.e. the direction of the dipole, depends on the electric field applied, and thus does not lead to a domain structure, as a typical ferroelectric does. This is evident in the comparison of first harmonic piezoresponse mappings shown in Fig. 4, including amplitude, phase, and frequency, acquired via dual amplitude resonance tracking (DART) technique [16] and processed via simple harmonic oscillator (SHO) model [14]. It is observed that Si exhibits rather uniform distribution of phase and frequency, which are largely featureless, while periodically poled LiNbO$_3$ reveals characteristic domain pattern in both phase and amplitude mappings. There is small amplitude variation observed in silicon, possibly due to spatial variation in surface condition.

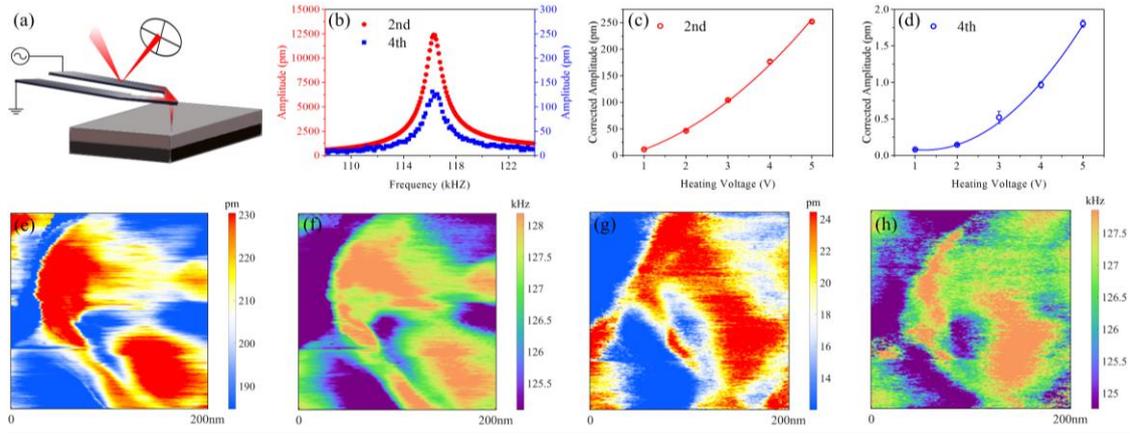

**Fig. 5** Si probed by STIM; (a) schematics of STIM; (b) 2$^{nd}$ and 4$^{th}$ harmonic resonant peak; (c) 2$^{nd}$ harmonic response versus heating voltage; (d) 4$^{th}$ harmonic response versus heating voltage; (ef) mappings of 2$^{nd}$ harmonic amplitude and frequency; (gh) mappings of 4$^{th}$ harmonic amplitude and frequency.

The electrochemical dipoles are presumably induced by ionic redistribution in Si, and if so, Vegard strain [19,20] may also arise as a result. To examine the ionic activities in Si, we use the recently developed STIM [11,12] to probe it, as schematically shown in Fig. 5a. An AC voltage was applied through the micro-fabricated probe with a localized resistive heater, inducing localized temperature fluctuation that is second harmonic to the heating voltage. This temperature fluctuation in turn induces local thermal expansion that is second harmonic to the heating as well, and the alternating thermal stress resulted induces ionic fluctuation and thus Vegard strain that turns out to be four harmonic to the heating [11,12]. Both second and fourth harmonic deflection of cantilever are captured in Fig. 5b near the cantilever-sample contact resonance of $\omega_0$, acquired separately under heating voltage frequency across $\omega_0/2$ and $\omega_0/4$, respectively. A large second harmonic thermal expansion peak that is universal in any material is observed in Fig. 5b, along with a fourth harmonic Vegard strain that confirms the ionic activities in silicon. The experiments were repeated under a number of different heating voltages, and after corrected by SHO model [14], the thermal and Vegard strain versus heating voltage are shown in Fig. 5cd, exhibiting functional variations that are approximately quadratic and quartic, as expected. The mappings of second harmonic thermal strain (Fig. 5e) and fourth harmonic Vegard strain (Fig. 5f) were acquired under DART [16], and they show small spatial variation, possibly due to the fluctuation of thermal probe during the scan that results in different cantilever deflection as seen. Nevertheless, the mappings of frequency $\omega_0$ (Fig. 5fh) are rather uniform and are consistent between second ((Fig. 5f)) and fourth (Fig. 5h) harmonic mappings, and thus the resonance tracking of DART is robust.

**Concluding Remarks**

By systematically investigating the first and second harmonic apparent piezoresponses in both vertical and lateral modes of SPM, we show that Si wafer exhibits apparent electromechanical response under a charged probe that may arise from induced electrochemical dipoles. As a result, the electromechanical response measured is dominated by the second harmonic quadratic response in vertical mode, and its polarity can be switched by the DC voltage with evolving coercive field and maximum amplitude. The ionic activity in silicon is also confirmed by STIM measurement.

**Acknowledgments**

We acknowledge National Key Research and Development Program of China (2016YFA0201001), US National Science Foundation (CBET-1435968), National Natural Science Foundation of China (11627801 and 11472236), the Leading Talents Program of Guangdong Province (2016LJ06C372), and Shenzhen Science and Technology Innovation Committee (KQJSCX20170331162214306).